\documentclass[]{spie}  

\usepackage{algorithm}
\usepackage{algpseudocode}
\usepackage{amsmath,amsfonts,amssymb}
\usepackage{graphicx}
\usepackage[colorlinks=true, allcolors=blue]{hyperref}
\usepackage{subcaption}
\usepackage{tikz}
\usetikzlibrary{patterns}
\usetikzlibrary{calc,decorations.markings,intersections,arrows.meta}
\usetikzlibrary{quotes,angles}

\newcommand{\DM}[1]{\textcolor{black}{#1}}

\title{A Multimodal Vision Sensor for Autonomous Driving}

\author[]{Dongming Sun}
\author[]{Xiao Huang}
\author[]{Kailun Yang}
\affil[]{Zhejiang University}

\authorinfo{Further author information: (Send correspondence to Dongming Sun)\\Dongming Sun: E-mail: dongmingsun@zju.edu.cn\\  Kailun Yang: E-mail: elnino@zju.edu.cn}

\begin{document} 
\maketitle

\begin{abstract}
This paper describes a multimodal vision sensor that integrates three types of cameras, including a stereo camera, a polarization camera and a panoramic camera. Each sensor provides a specific dimension of information: the stereo camera measures depth per pixel, the polarization obtains the degree of polarization, and the panoramic camera captures a $360^{\circ}$ landscape. Data fusion and advanced environment perception could be built upon the combination of sensors. Designed especially for autonomous driving, this vision sensor is shipped with a robust semantic segmentation network. In addition, we demonstrate how cross-modal enhancement could be achieved by registering the color image and the polarization image. An example of water hazard detection is given. To prove the multimodal vision sensor's compatibility with different devices, a brief runtime performance analysis is carried out.
\end{abstract}


\section{INTRODUCTION}
With the prosperity of autonomous driving, mobile robotics and drones, the demand for more integrated, accurate and robust sensors is rising apace. While there already exists quite a few prototypes of multimodal sensors designed for autonomous driving and smart robots, a large part of them are based on LiDAR systems\cite{varga2017super, goga2018fusing} and combinations of fisheye cameras to perceive the 3D surroundings\cite{yang2019can}, which results in cumbersome volume and non-negligible scanning time. In addition, those systems didn’t put emphasis on \DM{the} power consumption and require expensive computing resources if complex algorithms related to 3D point cloud modeling are involved. Moreover, the size and weight of existing multi-modal sensors might impose restrictions in possible applications scenarios, for instance, it could be an extreme burden for a drone to carry a LiDAR system with it.

In this paper, we present a novel multi-modal vision sensor which introduces some previously rarely considered equipment, including a polarization camera and a panoramic camera using Panoramic Annular Lens (PAL), together with a stereo camera to caputre large-scale 3D maps. We assemble the multi-modal vision sensor into a portable device to make it suitable to carry out experiments in outdoor environments. Equipped with the PAL, we are able to perceive the 360$^{\circ}$ environments. The polarization camera is used to detect specular materials such as glass, water, and metal, which could be potential dangers for \DM{autonomous} vehicles, robotics or drones. As for the stereo camera, it provides us with RGB images and depth measurements, which play essential roles in high-level vision perception \DM{tasks}.

To validate the versatility of our multi-modal vision sensor, several algorithms are developed and corresponding experiments are performed.
Firstly, the stereo camera is calibrated and a pair of left and right images are used to compute the depth maps.
Secondly, we design a semantic segmentation network based on a real-time convolutional model ERF-PSPNet\cite{yang2018unifying,yang2018semantic} with \DM{aggressive data augmentation} operations that \DM{have been demonstrated to be of critical relevance to achieve robustness in new, unseen domains\cite{yang2019can}.} \DM{The network is trained on a large-scale dataset like Cityscapes\cite{Cordts2016Cityscapes} and Mapillary Vistas\cite{neuhold2017mapillary}}. The \DM{yielded robust model} retrieves the left image from the stereo camera as input and outputs \DM{accurate} segmentation results \DM{of real-world scenes.}
Thirdly, we extract the pixels with high degree of linear polarization (DoLP) from the polarization camera images. By utilizing the dense depth \DM{information}, we view the left camera of the stereo camera (left camera for short) and the polarization camera as another cross-modal stereo pair and \DM{calibrate} their intrinsic and extrinsic parameters. In this way, we project pixels from the left camera image plane to that of the polarization camera thus attain depth \DM{measurements} for those high DoLP pixels, which \DM{are highly helpful to} avoid cars driving over slippery puddles on the highway, or feedbacking robots and drones not to crash into glass doors.
Furthermore, we undistort the \DM{panorama imaged by the Panoramic Annular Lens (PAL)} to have a more comprehensive perception of the surroundings.

By mounting the multi-modal vision sensor on top a car and crusing around the \DM{campus}, we prove the practicability of the sensor and the proposed algorithms. By comparing the frame rates on different computing devices such as PCs with modern GPUs, and portable propcessors like NVIDIA \DM{Jetson} TX2, we demonstrate that our multi-modal sensor with the embedded algorithms reaches excellent speed and reliability, making it exceptionally suitable for autonomous driving and robotic vision applications.

\section{Related Work}
Vision sensors are heavily applied in smart transportation \DM{systems} and autonomous robotics. A 6D-vision approach\cite{franke20056d} estimates 3D-position and 3D-motion for a large number of image points. Stereo vision and Kalman-\DM{filters} are employed to improve depth accuracy and compute the position and motion for each point.

Multimodal sensors have attracted huge attention for their ability to offer reliable information under complex and long-term scenarios since isolated sensors are \DM{prone to be} affected by the noisy environment. \citenum{stillman2001towards} presents a prototype system where different sensors \DM{have been} involved. A \DM{three-layer hierarchical} model to deal with multimodal fusion is proposed. The system network is able to obtain reliable state information about \DM{the} residents inside the home.

In \citenum{goga2018fusing}, semantic labeled images are fused with a LiDAR point cloud to provide good curb detection results in several shapes of roads. The usage of semantic information is to obtain ROI, latter fed into traditional \DM{LiDAR-based} methods which accelerate the computing speed. A LiDAR + monocular camera solution\cite{article} exploit both range and color information by projecting the 3-D point could of LiDAR onto the camera's frame to accomplish the road detection task. A conditional random field fusion method integrates the two road detection results from the image-based fully convolutional neural network and the point cloud based line scanning strategy.

\citenum{varga2017super} proposes a super sensor consisting of four fisheye cameras, four LiDARs and GPS/IMU to obtain 360-degree environment perception by harmonizing all the available sensor measurements. The point cloud is projected onto the semantic segmentation of images to get 3D segmentation results. Precise calibration and timestamp synchronization are crucial for such a complicated system.

\citenum{brooks2003tracking} implements a system of distributed sensors which work together to identify and track moving people using different sensing modalities in real time. A Kalman \DM{filter} is used to fuse information from cameras and laser scanners. Results of indoor person-tracking show that information from different kinds of sensors make each individual sensor more powerful.

\section{Multimodal Vision Sensor Design}
In this section, we \DM{present} the concept and design of the multimodal vision sensor.

\subsection{Stereo Camera}\label{setction3}
The stereo camera has been deemed as a competitive choice in scenarios where depth measurement is required due to its relatively \DM{compact} configuration and acceptable measurement accuracy. The procedure of stereo depth measuring normally consists of three steps\cite{kaehler2016learning}:
\begin{enumerate}
  \item stereo calibration
  \item stereo rectification
  \item stereo correspondence
\end{enumerate}
\par

The objective of stereo calibration is to find the rotation matrix $\mathbf{R}$ and translation vector $\vec{T}$ between the two cameras, such that for any three-dimensional point $\vec{P}$, its coordinates in left and right camera coordinates could be related as:
\begin{equation} \label{eq1}
\begin{split}
\vec{P_r} & = \mathbf{R} \cdot \vec{P_l} + \vec{T}\\
\end{split}
\end{equation}
A good enough estimation of $\mathbf{R}$ and $\vec{T}$ could be obtained using fifteen to thirty pairs of chessboard images.
\par
Stereo rectification makes stereo computation tractable by aligning image rows between the two cameras horizontally within a common image plane. This could be done by rotating the cameras twice respectively. First, rotate the left camera about its center of projection by $\mathbf{R}^{-\frac{1}{2}}$, and the right camera by $\mathbf{R}^{\frac{1}{2}}$, such that their principal rays end up parallel to each other. Next, an additional rotation $\mathbf{R}_{rect}$ puts the baseline parallel to the image planes, yielding the ideal stereo configuration\cite{loop1999computing}.
\par
Finally, we just need to run the stereo correspondence algorithm\cite{pollard1985pmf,scharstein2002taxonomy} to retrieve the depth as:
\begin{equation} \label{eq1}
\begin{split}
Z = \frac{f \cdot T}{x_r - x_l}
\end{split}
\end{equation}
where $f$ is the focal length and $T$ is \DM{the} stereo baseline.

\begin{figure}[h!]
  \centering
  \begin{subfigure}[b]{0.4\linewidth}
    \includegraphics[width=\linewidth]{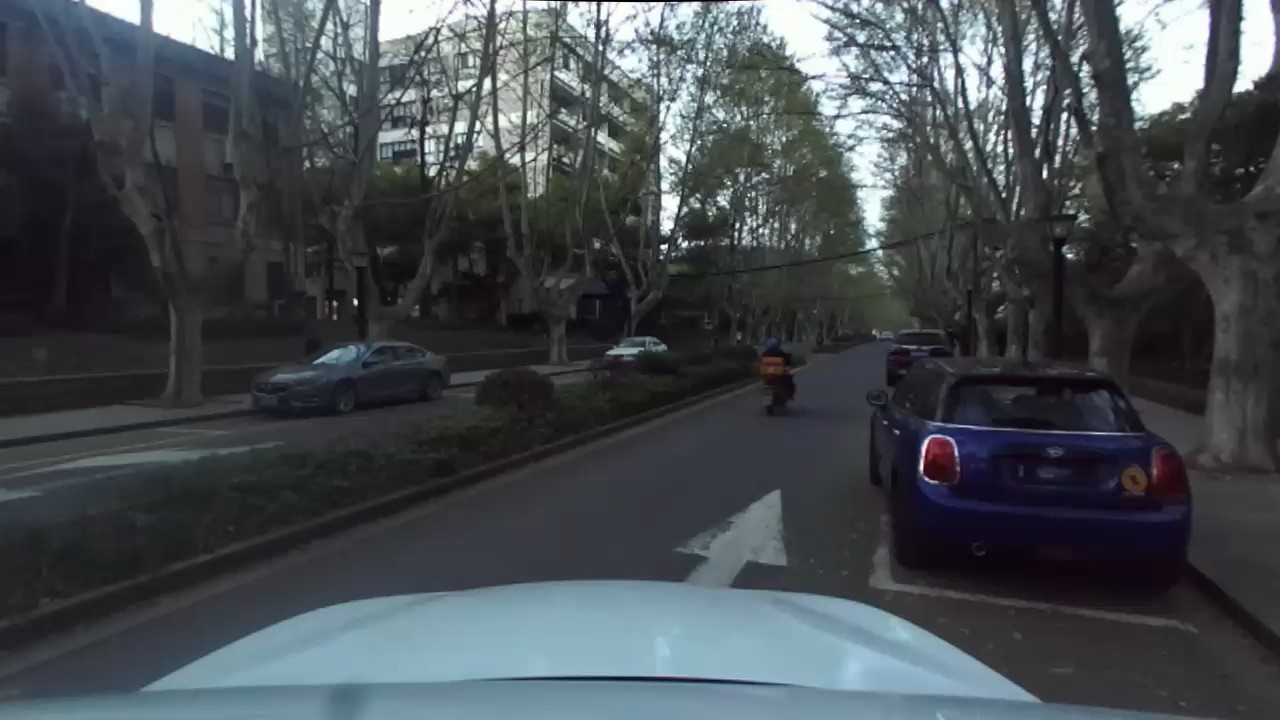}
    \caption{}
  \end{subfigure}
  \begin{subfigure}[b]{0.4\linewidth}
    \includegraphics[width=\linewidth]{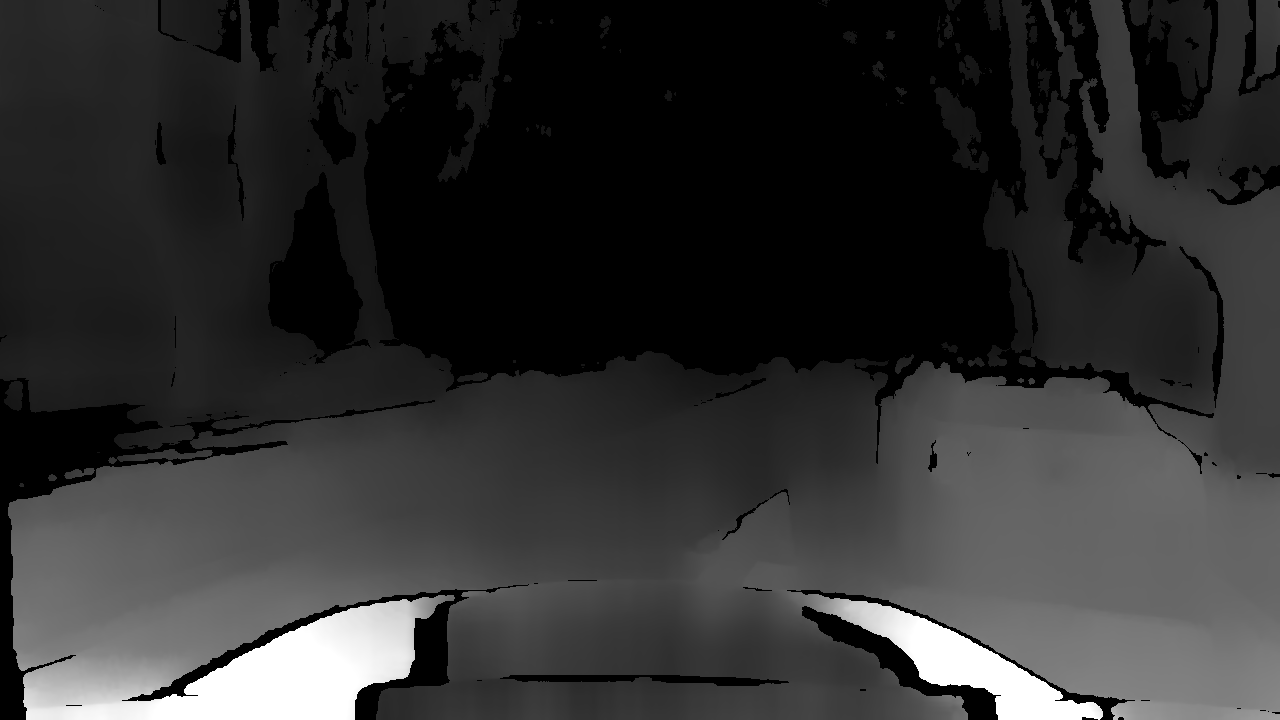}
    \caption{}
  \end{subfigure}
  \caption{Visualization of depth measurement: (a) the color image of the left camera; (b) the depth image aligned to the left color image.}
  \label{fig:color_and_depth}
\end{figure}

We choose ZED Mini as our stereo camera, which provides a $90^\circ(H)\times60^\circ(V)$ field of view and allows a depth range from 0.15m to 12m. Fig. \ref{fig:color_and_depth} visualises the depth measurement, where the depth image is aligned to the left color image.


\subsection{Polarization Camera}\label{section3.2}
It is known that when light strikes the surface between two mediums, both reflection and refraction occur\cite{huang2017target}. It is also know that any \DM{polarization} state can be resolved into a combination of two orthogonal linear polarizations.
The Fresnel's equation describes the amplitude ratio of perpendicular and parallel polarizations in both reflection and refraction waves:

\begin{equation} \label{eq2}
\begin{split}
r_s = \frac{n_1 \cos\theta_i - n_2\cos\theta_t}{n_1 \cos\theta_i + n_2\cos\theta_t}\\
t_s = \frac{2n_1\cos\theta_i}{n_1 \cos\theta_i + n_2\cos\theta_t}\\
r_p = \frac{n_2 \cos\theta_i - n_2\cos\theta_t}{n_2 \cos\theta_i + n_2\cos\theta_t}\\
t_p = \frac{2n_1\cos\theta_i}{n_2 \cos\theta_i + n_2\cos\theta_t}
\end{split}
\end{equation}

Here we use $s$ to refer to polarization of a wave's electric field perpendicular to the plane of incidence, and use $p$ to refer to polarization parallel to the plane of inncidence. The $r$ and $t$ denote the ratios of refraction and transmission. $n_1$ and $n_2$ are the refractive index of materials on two sides of the surface. We also assume an incidence angle of $\theta_i$ and a refractive angle of $\theta_r$. 

From Fresnel's equation, it can be deduced that different materials or surface roughness will pose impacts on the polarization of both reflection and refrection light, which could be a strong clue to distinguish certain types of objects\cite{rakovic1999light}. In polarmetric imaging, degree of linear polarization is a useful parameter describing the portion of an electromagnetic wave which is polarized, defined as $DOP = \frac{Power_{pol}}{Power_{total}}$. The Stokes parameters $\vec{S}=[S_0,S_1,S_2,S_3]^\intercal$ are a set of values that describe the polarization state in a mathematically convenient way. $S_0$ describes the total intensity of the optical beam; $S_1$ describes the preponderance of linear horizontally polarized light over linear vertically polarized light; $S_2$ describes the preponderance of linear $45^\circ$ polarized light over linear $135^\circ$ polarized light; and finally, the circular polarization component $S_3$ is negligible in natual scenes due to its slight amount. If given the Stokes parameters, one can solve the DOP as \DM{it is depicted in the following equation}\cite{goldstein2016polarized}:
\begin{equation} \label{eq3}
\begin{split}
DOP = \frac{\sqrt{S_1^2 + S_2^2}}{S_0}
\end{split}
\end{equation}

\begin{figure}[!h]
\centering
\begin{tikzpicture}
    \node[blue, align=center,fill=white] at (0.5,0.8){pixel};
    \draw[blue, very thick] (0,0.5) -- (1,0.5);
    \pgftransformxshift{1cm}{
     \draw[step=1cm, thick] (0,0) grid (4,4);
     \foreach \x in {0,2}
     \foreach \y in {0,2} {
     \draw[xshift=\x cm, yshift=\y cm, pattern=vertical lines] (0,1) rectangle (1,2);
     \draw[xshift=\x cm, yshift=\y cm, pattern=north east lines] (1,1) rectangle (2,2);
     \draw[xshift=\x cm, yshift=\y cm, pattern=north west lines] (0,0) rectangle (1,1);
     \draw[xshift=\x cm, yshift=\y cm, pattern=horizontal lines] (1,0) rectangle (2,1);
     \draw[red, step=2cm, very thick] (2,2) grid (4,4);
     }
     \draw[blue, step=1cm, very thick] (0,0) grid (1,1);
    }
     
     \pgftransformxshift{5cm}
     \pgftransformscale{2}{
     \draw[pattern=vertical lines] (0,1) rectangle (1,2);
     \draw[pattern=north east lines] (1,1) rectangle (2,2);
     \draw[pattern=north west lines] (0,0) rectangle (1,1);
     \draw[pattern=horizontal lines] (1,0) rectangle (2,1);
     \draw[step=1cm, thick] (0,0) grid (2,2);
     \draw[red, step=2cm, very thick] (0,0) grid (2,2);
     \node[text width=1cm,align=center,fill=white] at (0.5,1.5){$90^\circ$};
     \node[text width=1cm,align=center,fill=white] at (1.5,1.5){$45^\circ$};
     \node[text width=1cm,align=center,fill=white] at (0.5,0.5){$135^\circ$};
     \node[text width=1cm,align=center,fill=white] at (1.5,0.5){$0^\circ$};
     }
\end{tikzpicture}
\caption{The CMOS unit of a typical polarization image sensor.}
\label{cmos}
\end{figure}
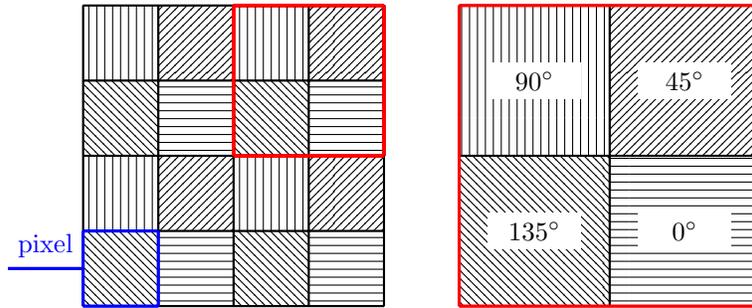

Modern polarized sensors incorporates a layer of polarizers above the photodiodes, where the polarizer array consists of four different angled wire-grid polarizers placed on each pixel as shown in Fig. \ref{cmos}. Therefore, Stokes parameters can be measured from the intensity of each pixel with Eq. \ref{polar_sensor}, from which we can further calculate the DOP for each unit.

\begin{equation} \label{polar_sensor}
\begin{split}
S_0 & = I_0 + I_{90} = I_{45} + I_{135}\\
S_1 & = I_0 - I_{90}\\
S_2 & = I_{45} - I_{135}
\end{split}
\end{equation}

\begin{figure}[h!]
  \centering
  \begin{subfigure}[b]{0.4\linewidth}
    \includegraphics[height=5cm]{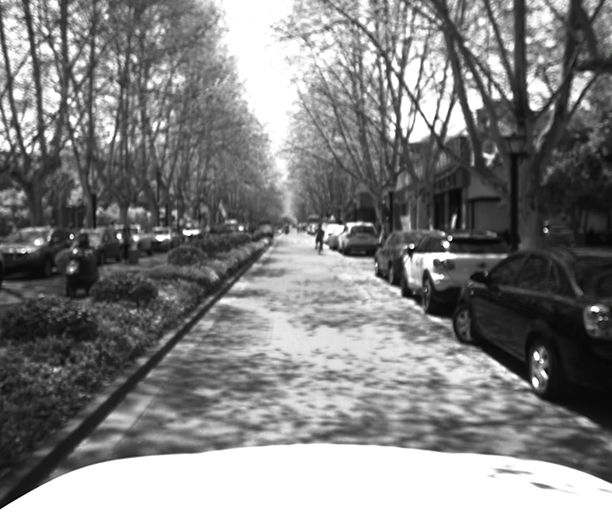}
    \caption{}
  \end{subfigure}
  \begin{subfigure}[b]{0.4\linewidth}
    \includegraphics[height=5cm]{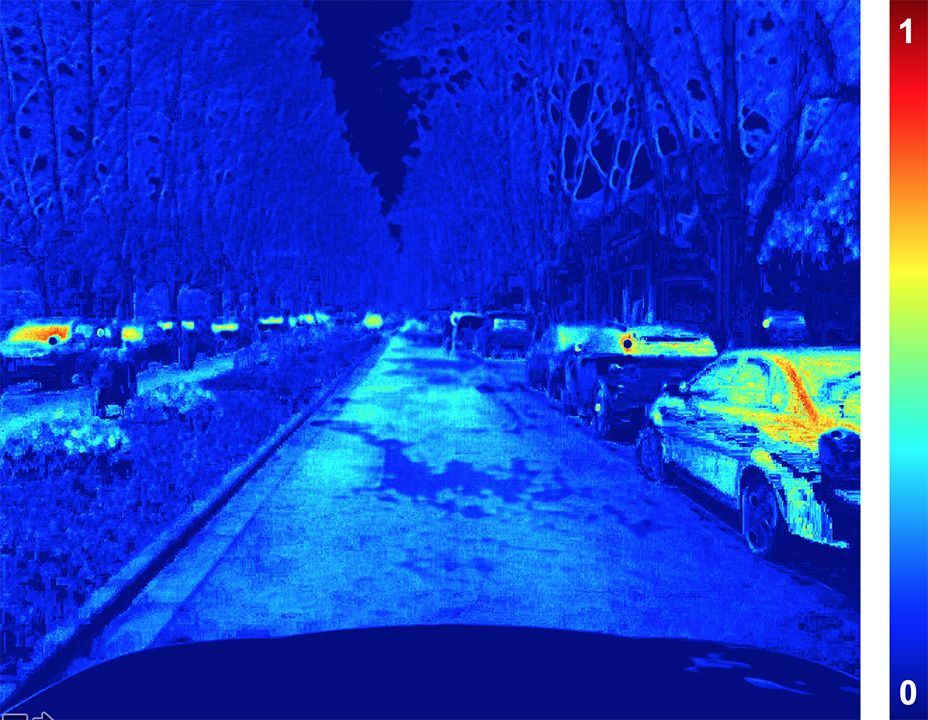}
    \caption{}
  \end{subfigure}
  \caption{Visualization of depth measurement. (a) the color image of the left camera; (b) the \DM{polarization} image aligned to the left color image.}
  \label{fig:dop}
\end{figure}

We choose the Lucid PHX050S C-mount camera and a VisionDatum LEM-7518CB-MP8 lens as our polarization sensor. The camera is equipped with a Sony IMX250MZR CMOS (Mono) chip whose polarizer configuration is the same as Fig. \ref{cmos}. A photo on a campus street is taken as shown in Fig. \ref{fig:dop}, in which the car windows reflect highly polarized light beams.

\subsection{Panoramic Camera}
\begin{figure}[!h]
\centering
\begin{tikzpicture}[
    ray/.style={decoration={markings,mark=at position .5 with {
      \arrow[>=latex]{>}}},postaction=decorate}
  ]
  \pgftransformscale{0.7} {

    \draw[very thick, name path=R0] (0,0) arc (-70:-110:2.6) coordinate (a);
    \draw[blue] (a) -- +(160:1) coordinate (a1);
    \draw[blue] (a) -- +(90:1) coordinate (a2);
    \draw (a) arc (100:140:3.5) coordinate (b);
    \draw[blue] (b) -- +(230:1) coordinate (b1);
    \draw[blue] (b) -- +(90:1) coordinate (b2);
    \draw pic[draw,fill=green!30,angle radius=0.3cm," $\theta_1$ " shift={(-2mm,3.5mm)}] {angle=a2--a--a1};
    \draw pic[draw,fill=green!30,angle radius=0.3cm," $\theta_2$ " shift={(-1.5mm,3.5mm)}] {angle=b2--b--b1};
    \draw[blue] (a) -- +(160:1) coordinate;
    \draw[blue] (a) -- +(90:1) coordinate;
    \draw[blue] (b) -- +(230:1) ;
    \draw[blue] (b) -- +(90:1);

    \draw let \p{b}=(b) in (b) -- (\x{b},\y{b}-30) coordinate (d);
    \draw [name path=R1] let \p{a}=(a), \p{b}=(b) in (\x{a}-\x{b}, \y{b}) coordinate (c) arc (40:80:3.5);
    \draw let \p{c}=(c) in (c) -- (\x{c},\y{c}-30) coordinate (e);
    \draw[very thick, name path=R2] (d) arc (215:255:4) coordinate (f);
    \draw[very thick] (e) arc (-35:-75:4) coordinate (g);
    \draw[name path=L0] (f) -- (g);

    \node (b) at ($(a)!0.5!(0,0)$) {};
    \draw[name path = Len] let \p{b}=(b) in (\x{b}, -5.5) ellipse (2 and 0.6); 

    \draw [very thick, name path=img] let \p{b}=(b) in (\x{b}-35, -9) -- (\x{b}+35, -9);

    \node[draw, circle, fill=black,scale=0.1, "Object point"] (o) at (3,1) {Object};
    \path[name path=line1] (o) -- (-5,-6.5);
    \draw[ray, thick,name intersections={of=R1 and line1,by={Int1}}] (o) -- (Int1);

    \path[name path=line2] (Int1) -- (-5,-4.5);
    \draw[ray, thick,name intersections={of=R2 and line2,by={Int2}}] (Int1) -- (Int2);

    \path[name path=line3] (Int2) -- +(1.65,4);
    \draw[ray, thick,name intersections={of=R0 and line3,by={Int3}}] (Int2) -- (Int3);
    
    \path[name path=line4] (Int3) -- +(1,-6);
    \draw[ray, thick,name intersections={of=L0 and line4,by={Int4}}] (Int3) -- (Int4);

    \path[name path=line5] (Int4) -- +(1.5,-6);
    \draw[ray, thick,name intersections={of=Len and line5,by={Int5}}] (Int4) -- (Int5);

    \path[name path=line6] (Int4) -- +(1,-6);
    \draw[thick,name intersections={of=Len and line6,by={Int6,Int7}}] (Int5) -- (Int7);

    \path[name path=line7] (Int7) -- +(0.1,-3);
    \draw[ray, thick,name intersections={of=img and line7,by={Int8}}] (Int7) -- (Int8);
  }
\end{tikzpicture}
\caption{Design principal of the panoramic \DM{annular} lens.}
\label{pal}
\end{figure}
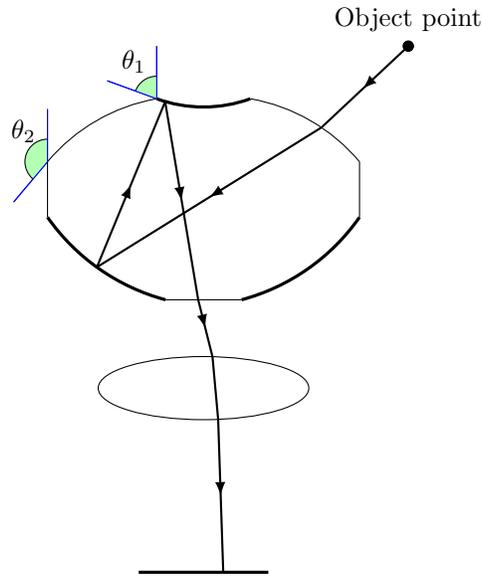

Panoramic Annular Lens (PAL) is a special kind of omnidirectional system with small size and extreme compactess\cite{nayar1997catadioptric}. The PAL is composed of a PAL block and one or more relay lens, as illustrated in Fig. \ref{pal}. A cylindrical field of view is projected to a planar annular image at a snapshot without the need to rotate the lens. Light rays in a vertical FOV from $\theta_1$ to $\theta_2$ are transformed to the image plane by two reflections and some refractions. The distortion can be well controlled due to the fact that the output rays are almost paraxial.

The PAL we design has a vertical FOV of $30^\circ \sim 95^\circ$. The focal length is 2.13mm, calculated via the f-theta law as $y'=f'\cdot\theta$. The relative aperature aperture is set to 1/3.2.

\begin{figure}[h!]
  \centering
  \includegraphics[width=\linewidth]{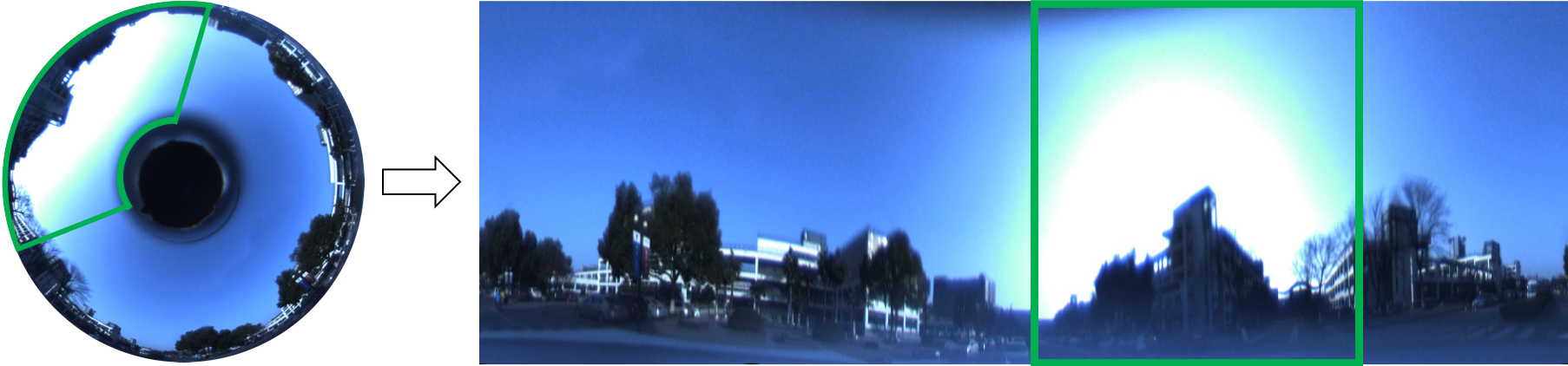}
  \caption{The unwrapping process.}
  \label{fig:unwrap}
\end{figure}

The \DM{acquired} PAL image is highly compact and \DM{not directly compatible} for human perception. \DM{For this reason}, we undistort and unwrap\cite{scaramuzza2006toolbox} the annular image for better visualization, shown in Fig. \ref{fig:unwrap}. It is simply done by unfolding the annulus and stitching the radiuses to form a rectangular image which is familiar to users.

\subsection{Sensor Integration}
We assemble the three cameras together to make a compact and portable multimodal vision sensor. It is designed to be mounted on top of the car and has room for placing an embedded computing device with additional power supply. Using this integrated sensor with a global shutter, we are able to collect multimodal data synchronically, facilitating advanced applications.

\begin{figure}[h!]
  \centering
  \includegraphics[width=\linewidth]{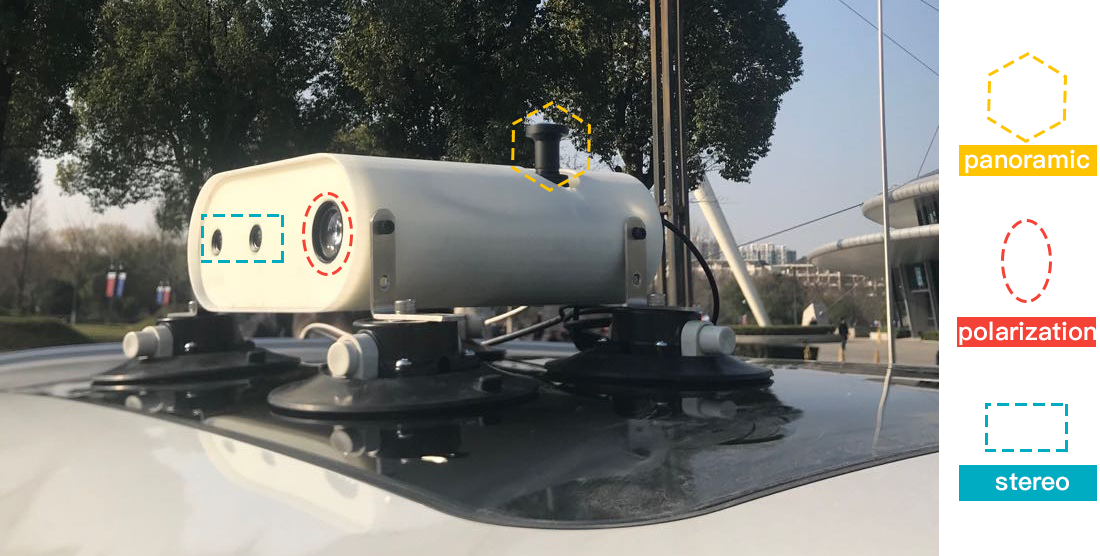}
  \caption{An overview of the integrated vision sensor.}
  \label{fig:assemble}
\end{figure}

\section{Information Processing and Data Fusion}
Until now, we've been treating the introduced sensors all alone. In this section, more information from the sensors will be utilized and more relevance among the sensors will be \DM{studied}. 

\subsection{Semantic Segmentation}\label{section4.1}
The stereo camera generates both color images and depth images. However, the depth information alone cannot empower complex tasks such as navigation. So it is necessary to understand what the camera is looking at and then the distance information might be valuable.
In autonomous driving, pixel-wise image segmentation has become essential on account of its ability to provide terrain awareness in a unified way\cite{yang2018unifying,yang2018semantic}. Specific types of objects like curbs, cars and pedestrians are of great importance to the backend control system. A number of CNN-based semantic segmentation networks have been proposed and they have different focuses. What we are concerned here is the tradeoff between the computational cost and the segmentation accuracy. Based on this criterion, we opt for the ERF-PSPNet proposed by\cite{yang2018unifying}. The ERF-PSPNet follows the encoder-decoder architecture, with an efficient non-bottleneck residual module in \DM{the} encoder to save computational resources and a pyramid pooling module in \DM{the} decoder to exploit more context both locally and globally.

\begin{figure}[h!]
  \centering
  \begin{subfigure}[b]{0.4\linewidth}
    \includegraphics[width=\linewidth]{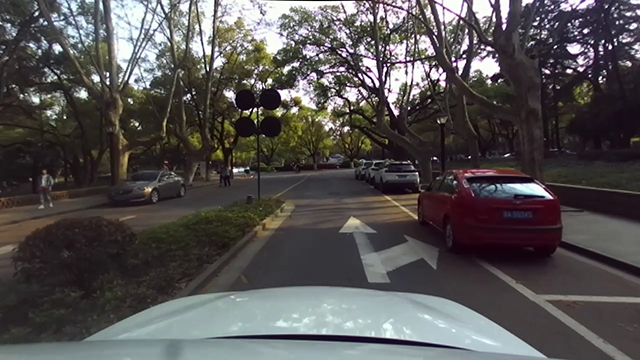}
    \caption{}
  \end{subfigure}
  \begin{subfigure}[b]{0.4\linewidth}
    \includegraphics[width=\linewidth]{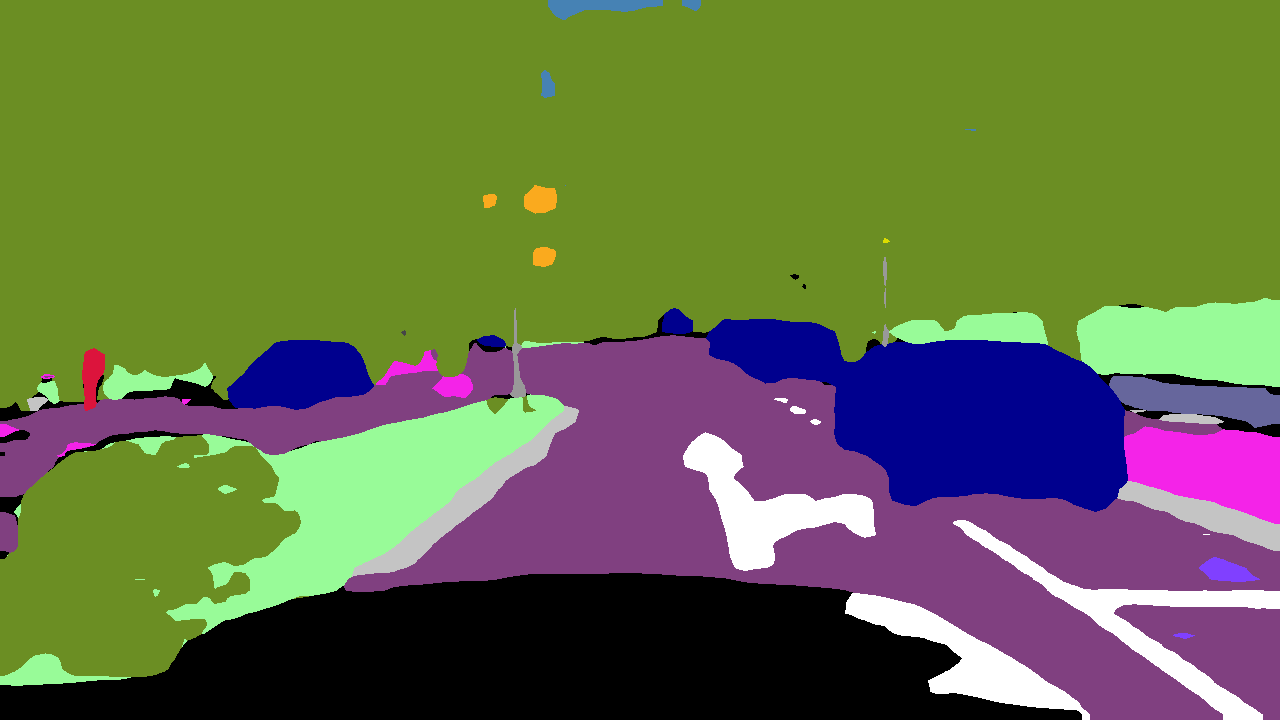}
    \caption{}
  \end{subfigure}
  \caption{Semantic segmentation of a campus street scene: (a) the color image of the left camera; (b) the \DM{semantic} segmentation.}
  \label{fig:semantic}
\end{figure}

In this work, we train the ERF-PSPNet on Cityscapes\cite{Cordts2016Cityscapes} and \DM{Mapillary Vistas\cite{neuhold2017mapillary}} datasets, and use the left color image from the stereo camera as input for inference. \DM{Fig. \ref{fig:semantic}} visualizes the segmentation result of a campus street scene.

The performance of our semantic segmentation network is also analyzed on the Cityscapes and \DM{Mapillary Vistas datasets} using the \DM{Intersection over Union} (IoU) as the metric. \DM{On the validation set of Cityscapes, we achieve 71.5\% (19 classes) of mean IoU (mIoU) at the resolution of 1024$\times$512. On Mapillary Vistas, we achieve a mIoU of 54.3\% for 27 classes, recorded in \textbf{Table} \ref{table:ss}.} The evaluation is done upon six classes which we reckon are important for autonomous driving. \DM{Training is defined in \cite{yang2019can}.}

\begin{table}[h!]
\centering
\begin{tabular}{|l|l|l|l|l|l|l|l|}
\hline
\textbf{Label} & mIoU & sky  & terrain & vegetation & sidewalk & person & car  \\ \hline
\textbf{IoU(\%)} & 54.3 & 98.0 & 65.2    & 88.6       & 68.5     & 64.6   & 88.4 \\ \hline
\end{tabular}
\caption{Semantic segmentation IoU of \DM{important} classes.}
\label{table:ss}
\end{table}

\subsection{Registration Between the Stereo Camera and Polarization Camera}
In \textbf{Section} \ref{section3.2} and \textbf{Section} \ref{section4.1}, the advantages of semantic segmentation and polarization perception are demonstrated separately. To some extent, they're complementary to each other. One example is the water hazard detection. Limited to the abundance of training dataset, there are some categories we are concerned about but missing in the training data. The Cityscapes \DM{dataset}, for instance, doesn't annotate the water areas which hardly appear in its collection. However, slippery ground can become a threat to \DM{autonomous} cars. While annotating extra categories and retraining the network is tedious and causes heavy work, the polarization camera \DM{is very beneficial for perceiving specular areas, significantly reducing the burden.} \DM{This is due to that} in natural scenes, the light rays reflected by the water is highly polarized thus could be extracted by the polarization camera. In this case, it is the physical characteristic of the target of interest that accomplishes the detection task, which is highly efficient and accurate.

But still, we may want all the segmentation results including the water hazards combined together so that more robust decisions could be made by \DM{upstream} navigation algorithms. Considering that the input of the semantic segmentation network is the left color image of the stereo camera, we have to register the polarized image to the left color image. In \textbf{Section} \ref{setction3}, the stereo calibration is introduced. Similarly, the polarization camera and the left stereo camera could be treated as another stereo pair. 

\begin{figure}[h!]
  \centering
  \begin{subfigure}[b]{0.3\linewidth}
    \includegraphics[width=\linewidth]{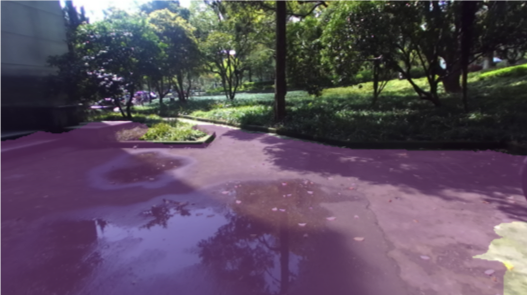}
    \caption{}
  \end{subfigure}
  \begin{subfigure}[b]{0.2\linewidth}
    \includegraphics[width=\linewidth]{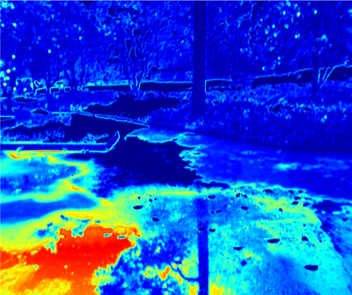}
    \caption{}
  \end{subfigure}
  \begin{subfigure}[b]{0.3\linewidth}
    \includegraphics[width=\linewidth]{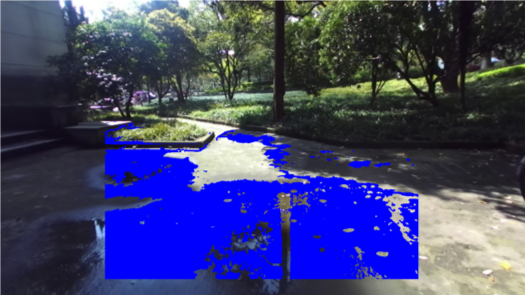}
    \caption{}
  \end{subfigure}
  \caption{Water hazard detection. (a) semantic segmentation using the left color image of the stereo camera, where the ground area is highlighted; (b) the pseudo-color DOP measurement retrieved from the polarization camera; (c) the blue area highlights the water hazard detected after registrating and fusing (a) and (b).}
  \label{fig:water}
\end{figure}

Let $\mathbf{u}^P_{color}= (u^P_{color},v^P_{color})$ denote a spatial point $P$'s coordinate on the left color image, then the goal is to find out the point's coordinate on the polarization image $\mathbf{u}^P_{polar}=(u^P_{polar},v^P_{polar})$. Suppose after calibration, we have both cameras' intrinsics and extrinsics, the registration process could be done by:
\begin{equation} \label{eq4}
  \mathbf{u}^P_{polar} = \pmb{\pi}(\mathbf{K}_{polar} \mathbf{T} \mathbf{K}_{color}^{-1} (z\cdot \dot{\mathbf{u}}^P_{color}) ) 
\end{equation}
where
\begin{align} \label{eq5}
\begin{split}
&\mathbf{T}=
\begin{bmatrix}
\mathbf{R} & \vec{T}\\
\vec{0} & 1
\end{bmatrix}\\
&\dot{\mathbf{u}} = (u,v,1)^\intercal\\
&\pmb{\pi}(\mathbf{p}) = (x/z,y/z) \text{ for } \mathbf{p} = (x,y,z)
\end{split}
\end{align}
and $\mathbf{K}$ denote the intrinsics of the camera.

To find out the water hazard area, Algorithm~\ref{waterdetection} inspired by \cite{yang2018perception} is applied, and the segmentation result is shown in Fig.~\ref{fig:water}. Note that the FOV of the polarization camera is smaller \DM{than} that of the stereo camera, so the polarization image only covers part of the color image.

\begin{algorithm}[h!]
  \caption{Water hazard detection. }\label{waterdetection}
  \begin{algorithmic}[1]
    \For{$\mathbf{u}_{color}\in\Omega_{color}$}\Comment{Suppose the semantic segmentation is done before the loop}
      \State {$ \delta \gets 0.6$}  \Comment {input water area DOP threshold}
      \State \texttt{compute corresponding} $\mathbf{u}_{polar}$ \texttt{according to \textbf{Eq.}} \ref{eq4}
      \If {$\mathbf{u}_{polar}\in\Omega_{polar}$}
          \If {$class(\mathbf{u}_{color})=$ road \textbf{and} ($DOP(\mathbf{u}_{polar})\ge\delta)$}
              \State $class(\mathbf{u}_{color})\gets$ \texttt{water hazard}
          \EndIf
      \EndIf
    \EndFor
  \end{algorithmic}
\end{algorithm}

\subsection{Runtime Performance Analysis}
The multimodal sensor we design could be applied to different computing devices. To validate its runtime performance, we test the \DM{Frames Per Second} (FPS) using different configurations on different platforms, and the results are collected in \textbf{Table} \ref{table:runtime}. The items that we run during the test include color images, depth images, semantic segmentation, polarization perception and panoramic images. Note that the resolution is applied to both the stereo camera and the polarization camera. \DM{The results show that our proposed multimodal can produce a rich set of raw/high-level visual data for upper-level autonomous driving applications in near real time, but the speed could be further optimized.}

\begin{table}[h!]
\centering
\begin{tabular}{lll}
\hline
Platform                & Resolution     & FPS  \\ \hline
Nvidia TX2              & $320\times240$ & 5.3  \\ 
Nvidia TX2              & $640\times480$ & 2.8  \\ 
1080Ti \& Intel i5-8400 & $320\times240$ & 21.2 \\ 
1080Ti \& Intel i5-8400 & $640\times480$ & 11.5 \\ \hline
\end{tabular}
\caption{Runtime performance on different platforms.}
\label{table:runtime}
\end{table}

\section{Conclusions and Future Work}
A multimodal vision sensor is built by integrating a stereo camera, a polarization camera and a panoramic camera. We demonstrate how to measure depth and polarization. The unwrapping process of the panoramic image is also shown in detail. The advantages of combining depth measurement and polarization measurement \DM{are} demonstrated.

We show how cross-modal registration could be done by calibration and reprojection between different cameras. An example of water hazard detection is given to explain how the fusion of color, depth and polarization information could help improve safety in autonomous driving. Furthermore, experiments are carried out to prove the sensor's compatibility with different computing platforms from portable devices to PCs.

The proposed multimodal sensor has already been used in diverse intelligent vehicles systems for panoramic scene parsing\cite{yang2019can}, visual topological localization\cite{cheng2019panoramic} and nighttime semantic understanding\cite{romera2019bridging,sun2019see}. In the future, we aim to optimize the speed and deploy our sensor in more transportation applications.

\section*{Acknowledgment}

This work has been partially funded through the project ``Research on Vision Sensor Technology Fusing Multidimensional Parameters'' (111303-I21805) by Hangzhou SurImage Technology Co., Ltd. 

\bibliography{report} 
\bibliographystyle{spiebib} 

\end{document}